%% file: main.tex
\documentclass[conference]{IEEEtran}
\usepackage{color}
\usepackage{amsmath,amssymb,amsfonts}
\usepackage{algorithmic}
\usepackage{graphicx}
\usepackage{textcomp}
\usepackage{booktabs}
\usepackage{xcolor}
\usepackage{listings}
\graphicspath{{./figures/}}

\ifCLASSINFOpdf
\else
\fi
\definecolor{verylightgray}{rgb}{.97,.97,.97}

\lstdefinelanguage{Solidity}{
	keywords=[1]{anonymous, assembly, assert, balance, break, call, callcode, case, catch, class, constant, continue, contract, debugger, default, delegatecall, delete, do, else, event, export, external, false, finally, for, function, gas, if, implements, import, in, indexed, instanceof, interface, internal, is, length, library, log0, log1, log2, log3, log4, memory, modifier, new, payable, pragma, private, protected, public, pure, push, require, return, returns, revert, selfdestruct, send, storage, struct, suicide, super, switch, then, this, throw, transfer, true, try, typeof, using, value, view, while, with, addmod, ecrecover, keccak256, mulmod, ripemd160, sha256, sha3}, 
	keywordstyle=[1]\color{blue}\bfseries,
	keywords=[2]{address, bool, byte, bytes, bytes1, bytes2, bytes3, bytes4, bytes5, bytes6, bytes7, bytes8, bytes9, bytes10, bytes11, bytes12, bytes13, bytes14, bytes15, bytes16, bytes17, bytes18, bytes19, bytes20, bytes21, bytes22, bytes23, bytes24, bytes25, bytes26, bytes27, bytes28, bytes29, bytes30, bytes31, bytes32, enum, int, int8, int16, int24, int32, int40, int48, int56, int64, int72, int80, int88, int96, int104, int112, int120, int128, int136, int144, int152, int160, int168, int176, int184, int192, int200, int208, int216, int224, int232, int240, int248, int256, mapping, string, uint, uint8, uint16, uint24, uint32, uint40, uint48, uint56, uint64, uint72, uint80, uint88, uint96, uint104, uint112, uint120, uint128, uint136, uint144, uint152, uint160, uint168, uint176, uint184, uint192, uint200, uint208, uint216, uint224, uint232, uint240, uint248, uint256, var, void, ether, finney, szabo, wei, days, hours, minutes, seconds, weeks, years},	
	keywordstyle=[2]\color{teal}\bfseries,
	keywords=[3]{block, blockhash, coinbase, difficulty, gaslimit, number, timestamp, msg, data, gas, sender, sig, value, now, tx, gasprice, origin},	
	keywordstyle=[3]\color{violet}\bfseries,
	identifierstyle=\color{black},
	sensitive=false,
	comment=[l]{//},
	morecomment=[s]{/*}{*/},
	commentstyle=\color{gray}\ttfamily,
	stringstyle=\color{red}\ttfamily,
	morestring=[b]',
	morestring=[b]"
}

\begin{document}
%
\title{Towards Efficient Integration of Blockchain for IoT Security: The Case Study of IoT Remote Access}

\author{Anonymous Submission}

\author{
	
	\IEEEauthorblockN{Chenglong Fu}
	\IEEEauthorblockA{Temple University\\ Email: chenglong.fu@temple.edu}
	\and
	\IEEEauthorblockN{Qiang Zeng}
	\IEEEauthorblockA{University of South Carolina\\Email: zeng1@cse.sc.edu}
	\and
	\IEEEauthorblockN{Xiaojiang Du}
	\IEEEauthorblockA{Temple University\\ Email: xjdu@temple.edu}	
}



%

\IEEEoverridecommandlockouts

\maketitle

\begin{abstract}
The booming Internet of Things (IoT) market has drawn tremendous interest from cyber attackers. The centralized cloud-based IoT service architecture has serious limitations in terms of security, availability, and scalability, and is subject to single points of failure (SPOF). Recently, accommodating IoT services on blockchains has become a trend for better security, privacy, and reliability. However, blockchain's shortcomings of high cost, low throughput and long latency make it unsuitable for IoT applications. In this paper, we take a retrospection of existing blockchain-based IoT solutions and propose a framework for efficient blockchain and IoT integration. Following the framework, we design a novel blockchain-assisted decentralized IoT remote accessing system, \textsc{RS-IoT}, which has the advantage of defending IoT devices against zero-day attacks without relying on any trusted third-party. By introducing incentives and penalties enforced by smart contracts, our work enables ``\emph{an economic approach to cybersecurity,}'' thwarting the majority of attackers who aim to achieve monetary gains. Our work presents an example of how blockchain can be used to ensure the fairness of service trading in a decentralized environment and punish misbehaviors objectively. We show the security of RS-IoT  via detailed security analyses. Finally, we demonstrate its scalability, efficiency, and usability through a proof-of-concept implementation on the Ethereum testnet blockchain.\footnote{An earlier version of this paper was submitted to ACM CCS’18 on May 9th, 2018. This version contains some minor modifications based on that submission.}
\end{abstract}


%




\input{./sections/introduction2}

\input{./sections/background}
\input{./sections/investigation}
\input{./sections/casestudy}

\input{./sections/mydesign}

\input{./sections/proofofdelivery}

\input{./sections/penalty}

\input{./sections/analysis}

\input{./sections/experiment}
\input{./sections/relatedwork}
\input{./sections/conclusion}

\bibliographystyle{IEEEtranS}
\bibliography{cite}

\end{document}

%% file: sections/introduction2.tex
\section{Introduction}\label{sec:into}
The IoT market is flourishing. According to Gartner's report in 2018~\cite{predict2018gartner}, there is predicted to have 14.2 billion connected things in use in 2019 and 25 billion by 2021.  However, security concerns are raised along with the growth of the IoT market. A series of horrible IoT-related attacks have been seen during the past years such as the Mirai botnet attack~\cite{Antonakakis2017}, BrickerBot attack~\cite{kolias2017ddos}, Deutsch Telekom TR-069 attack~\cite{Antonakakis2017}. Newly developed fancy pwns and hacks targeting IoT devices like~\cite{pierre2017dlink,awesome-iot-attacks} are emerging every day. Considering the scale of IoT devices, securing them becomes a non-trivial task. For example, the Mirai botnet attacker launches the record-breaking DDoS attack by recruiting more than 600K compromised IoT devices as his bot army, causing inaccessibility of many high-profile websites such as Twitter, Reddit, and Netflix.

Securing these vulnerable IoT devices is challenging. Not only because those low-cost IoT devices are lacking of computing resources and I/O peripherals, but also due to the IoT vendors' lax on implementing secure softwares. Many manufacturers are busy rolling out products with novel features while leaving security flaws unpatched for years~\cite{dlinknopatch,awesome-iot-attacks,tschofenig2017report,schneier2014internet}. Given IoT devices' actuality of insecurity, it is reasonable to have the assumption of \emph{access-to-attack}, with which any attacker that has direct access to the IoT device's open port is supposed being able to take down the device. Modern IoT vendors try to address this problem by anchoring the access entries of their products on endpoint instances hosted by cloud servers~\cite{alrawi2019sok}. They migrate critical services such as authentication, remote administration, and data collection from vulnerable IoT devices to the more secure cloud server. Unfortunately, the insecurity of cloud servers is never a piece of news with prominent examples of high-profile compromises including Equifax~\cite{bernard2017equifax}, Dropbox, and the US voter database~\cite{walters2014cyber}.  Since the Cloud endpoint serves as the entry and is trusted by a large number of IoT devices, it may, on the contrary, give adversaries an additional arsenal for launching large scale attacks~\cite{zhang2015cloud,singh2016twenty}.

Recent advancements on the blockchain and smart contracts inspire researchers to seek blockchain-based solutions for its intrinsic advantages on decentralization, faulty tolerance, and data integrity. However, incorporating blockchain and IoT is non-trivial due to the blockchain's  characteristics and the IoT's requirements. On the one hand, billions of IoT devices are running 24/7 and produce enormous amounts of data to be timely stored and processed. On the other hand, blockchain usually has limited throughput and is costly. Although the smart contract theoretically achieves Turing-complete, the  Proof-of-Work (PoW) consensus mechanism makes it not only expensive but also slow.

Currently, most research on blockchain-based IoT solutions are still using the blockchain as a substitution of the cloud server. Their approaches to solving the aforementioned challenges can be roughly categorized into three types: 1) They choose to study specific services that are latency insensitive and bring low overhead, for example, IoT authentication and identity management. 2) They use private or permissioned blockchains instead of public blockchains to avoid cost and throughput issues. 3) They turn to edge computing where edge servers are introduced to mitigate the IoT's resource constrains for interacting with the blockchain. These workarounds are either making a trade-off between the usability and security or are limited on specific tasks. A general framework to efficiently integrate blockchain into IoT systems is still an open question.

\paragraph{This Work}
In this paper, we try to fill this gap by rethinking the blockchain's role in the IoT service architecture. Instead of being the host to accommodate services directly, blockchain is more suitable to be a service trading platform where service users and third-party providers can discover each other, establish commission relationships, and settle service fees. The service itself is undertaken by independent third-party providers.  The participation of third-party providers  decouples the relationship between IoT devices and vendor operated cloud servers, which allows IoT devices switch to any provider for better service security and quality. To realize it, the following questions need to be answered:

\begin{itemize}
	\item[\textbf{Q1:}] How to motivate the participation of third-party service providers?
	\item[\textbf{Q2:}] How to establish a mutual trust between service users and providers?
	\item[\textbf{Q3:}] How to prevent malicious behaviors like cheating, attacking, and denial of service?
\end{itemize}

We propose \textsc{RS-IoT}, a novel blockchain assisted IoT relay sharing system as a case study of the IoT remote access service. In the system, we introduce third-party relay servers to substitute the centralized message broker~\cite{happ2017meeting} for enabling two-way relayed communications between a user's controller client device and an IoT device. IoT device owners on this platform can freely commission any relay servers for their devices instead of using those designated by device vendors. First, the decentralized nature of the proposed technique resolves the SPOF and scalability issues. Leveraging the power of the smart contract~\cite{buterin2014next}, we design a transparent, self-governing relay service trading protocol where \emph{monetary incentives} are used to motivate third-party relay providers' participation and deter potential malicious behaviors. Furthermore, fair and objective disputation arbitration and attack handling are achieved without any trusted authorities by using our proof-of-delivery scheme which involves off-chain proof generation and on-chain verification. As a result, third-party relay servers get the incentive to shield their customer IoT devices, which gives vulnerable IoT devices additional protection against zero-day attacks.







\paragraph{Contribution}
We rethink the integration of blockchain for IoT systems after a comprehensive literature investigation. Based on what, we propose a novel blockchain-assisted decentralized relay sharing system as a solution to the IoT remote accessing problem. Our contributions are summarized as follows:
\begin{itemize}
	\item We propose a practical framework for Blockchain enabled IoT services and provide the guideline to resolve its inherent shortcomings of cost, throughput, and latency.  
	
	\item We propose the proposed \textsc{RS-IoT}, which to the best of our knowledge, is the first decentralized relay architecture designed for IoT remote access.
	
	\item We design a smart contract based relay service trading system where disputes are resolved by using smart contracts.

	\item We achieve precautionary defense against future unknown malware by presenting the misbehavior reporting scheme which is inspired by the concept of N-version programming \cite{chen1995n}.  Malicious attacking behaviors are deterred because once the malware attack fails on \emph{any} IoT device, the attacker's deposit will be confiscated to cause direct financial loss. 
	
\end{itemize}

The rest of the paper is organized as follows. We present the related background knowledge in Section~\ref{sec:background}. In Section~\ref{sec:integration}, we review existing researches in terms of blockchain and IoT integration and propose our framework for converting cloud-based IoT services to blockchain-assisted distributed services. In Section~\ref{sec:case}, we study the use case of IoT remote access and give the threat model. The design overview  of RS-IoT is illustrated in~\ref{sec:design}. After that, Section~\ref{sec:pod} and Section~\ref{sec:penalty} describe the details of proof-of-delivery and malicious behavior reporting. Security analysis and experiments are presented in Section \ref{sec:secanalysis} and Section \ref{sec:exp}, respectively. We survey related works in Section~\ref{sec:relatedwork} and have discussions of some implementation issues in Section~\ref{sec:dis}. Finally, we conclude the paper in Section~\ref{sec:con}.

%% file: sections/background.tex
\section{Background}\label{sec:background}
\subsection{N-version Programming}\label{sec:nvp}
N-version programming (NVP) \cite{chen1995n} is a method used in software engineering in which multiple versions of software are created from the same copy of initial specifications. These equivalent copies of softwares are developed independently in different approaches. It is proved that NVP can greatly reduce the influence from identical software faults. The concept has already been used as an effective defense method against software flaws \cite{cox2006n}. For our proposed relay sharing system, the variety of IoT devices' software implementation makes it impossible for attackers to launch universally applicable attacks and imposes risks when they make unsuccessful attempts.


\subsection{Non-Repudiable TLS}
TLS-N \cite{ritzdorf2018tls} is an extension of the current TLS protocol. It adds non-repudiation features. Traditional TLS/SSL protocol only verifies the identity of the other end at the beginning of the session and assumes the consistency of identity after the encrypted session is established. Although it is effective to defeat the man-in-the-middle attack and impersonation attack, it does not support communication forensics with non-deniable proofs of packets sending. Unlike the normal TLS protocol which only uses HMAC in application data packets for message integrity checking, TLS-N enables signatures on each packet transmitted between the communicating peers. By adding a verifiable signature created with the private key, the sender cannot deny sending certain content. The TLS-N protocol timely fills the gap between the imperfection of traditional TLS and the requirements of objective and verifiable message publication for smart contracts' execution. It is built on the base of normal TLS protocol and proved to be able to generate proofs that are not only non-repudiable but also privacy-preserving. As tested by the author, TLS-N only incurs less than 1.5 milliseconds increase of time overhead for each HTTP request compared to the original version of TLS protocol and costs no more than 3 USD for verifying proofs on the public Ethereum blockchain. In our work, TLS-N signatures of malicious packets are used by our reporting system as the evidence of relay servers' misbehaviors.

\subsection{Smart Contract}
Smart contract~\cite{buterin2014next} consists of pieces of script code on blockchain to deal with digital assets. It is executed by all miner nodes in the isolated virtual environment. The accepted execution results are recorded on the decentralized ledger through consensus mechanisms which ensures it to be trustworthy and tampering-resistant. Smart contracts provide two types of accounts: the personal account is owned by the participating node and protected by the private key; the contract account points to the instance of smart contract code without private keys. Both types of accounts have the account addresses and are able to hold digital assets.

%% file: sections/investigation.tex
\section{Blockchain and IoT Integration}\label{sec:integration}
Cloud platforms have been proved to help enhance resource constraint IoT devices against security threats~\cite{sajid2016cloud,parwekar2011internet,srivastava2015secure}. However, the risk of single point of failure and the problem of scalability bottleneck entice IoT vendors to shift their services from the centralized architecture to the decentralized form. The blockchain's intrinsic natures of decentralization, tamper-resistance, and autonomy make it a promising candidate to be used in the IoT paradigm. Furthermore, the emergence of smart contracts provides a practical way for Turning-complete trusted global computer, which inspires the proposal of autonomous IoT system~\cite{brody2014device}. 

However, these wonderful security benefits of the blockchain are not to be taken for granted. The `world computer' is achieved by using numerous blockchain nodes as redundant backups. This brings high cost and latency for operations on public blockchains. Moreover, despite blockchain's good scalability of accommodating unlimited participating nodes, the whole network throughput of transaction processing is limited. For instance, Bitcoin has a constant throughput of 10 minutes per block, which is equivalent to 7 transactions per second and Ethereum allows 15 seconds per block. Another issue is cost. All operations that modify the public blockchain's state are subject to transaction fees. Even simple operations like storing or changing a byte in the blockchain can be expensive, let alone complicated tasks such as data processing. Considering the IoT's requirement of large scale and low cost, the use of blockchain becomes unpractical.

To find a general and viable solution, we first review the existing research works of blockchain-based IoT security mechanisms and analyze their strengths and weaknesses on addressing the two mentioned constraints. Based on the retrospection, we propose our efficient integration framework.


\subsection{Retrospection of Existing Works}
Since the proposal of smart contracts in 2014, there have been many research works regarding the paradigm of IoT and blockchain integration. We choose three well-presented survey papers~\cite{panarello2018blockchain,christidis2016blockchains,conoscenti2016blockchain} as indices to collect notable works related to this topic for further analyses and discussions. In contrast to these survey papers, we focus on  categorizing and evaluating the technical methods used by these papers to solve the aforementioned problems. 
Specifically, we evaluate them in terms of 6 criteria: 

\begin{itemize}
	\item \textbf{Use Case:} The target application or service the work aims to provide for IoT.
	\item \textbf{Service Architecture:} Participating entities and their roles in service.
	\item \textbf{Service Requirements:} The requirements on throughput, latency and cost.
	\item \textbf{Scalability:} The scale of deployment.
	\item \textbf{Blockchain Specs:} The type of blockchain and consensus algorithm they use.
	\item \textbf{Attack Resilience:} Whether their system can exclude the malicious nodes and recover from a system failure.
\end{itemize}

Based on these criteria, we categorize the collected papers into the following two types:
\paragraph{Blockchain as a Ledger}
In this type, the blockchain is used as an immutable ledger or a distributed database to store critical information. This is the most straightforward way for the integration of IoT and blockchain and is adopted by many early works. \cite{wu2018out} proposes to use the blockchain to store and deliver out-of-band authentication credentials. 
Since authentication only occurs when devices request sensitive information from the cloud server, it has low requirements for throughput and latency. The authors indicate that they use the Eris blockchain, but they do not specify any details about the implementation, nor a solution to reduce the involved transaction fees. \cite{dorri2017blockchain} designs a multi-layer architecture for blockchain powered smart home access control. It employs a centralized device to process all transactions and generate new blocks to get rid of the overhead brought by the Proof-of-Work. However, this design contradicts the core concept of decentralization of blockchain and undermines the security. Some other works straightforwardly leverage the blockchain's immutable feature to facilitate distributed data storage and sharing as discussed in \cite{zyskind2015decentralizing} and \cite{worner2014your}. They use public blockchain to store either raw data or the hash of data. These works avoid the cost and throughput constraints by either choosing specific services that have low requirements for transaction frequency and latency or using a single centralized miner.




\paragraph{Blockchain as a Service}
Inspired by the concept of decentralized applications~\cite{raval2016decentralized}, some works regard the smart contract as a trusted computing platform and build more applications on it. The first category is decentralized PKIs based on the smart contract.  \cite{chen2018certchain} introduces an additional party named `bookkeeper' to form the backbone of a permissioned blockchain and store the certificate revocation list. But it does not specify the origin and the incentive of `bookkeepers', which implies a limited number of available bookkeepers. Certcoin~\cite{fromknecht2014certcoin} chooses to build a decentralized trust system based on Namecoin, a public blockchain for DNS services. Although the author finds a solution to mitigate end users' storage burdens, all certificates and public keys are still held on the blockchain. Access control and authorization is another hot topic for blockchain and IoT integration. `WAVE'~\cite{andersen2017wave} proposes a city-wide blockchain to store the metadata of permissions for supporting a variety of access control patterns. The authors craft a set of smart contracts to automatically handle complicated out-of-order access permission delegations. A bunch of other IoT authorization solutions~\cite{ourad2018using,alexopoulos2018towards,stanciu2017blockchain} basically use the similar architecture that employs smart contracts to automatically enforce access control policies. They try to address the cost and throughput issues either by using permissioned blockchains~\cite{chen2018certchain,stanciu2017blockchain,alexopoulos2018towards}, or delicately designing the contract code to minimize the frequency of modifying the global state of blockchain~\cite{ourad2018using,andersen2017wave}.

In summary, the aforementioned works basically follow two approaches. First, they choose specific services that require a low frequency of issuing transactions to avoid intensive on-chain operations which are subject to fees and latencies. Reading the blockchain data would not cause a state change and has no cost or minimal overhead. Therefore, the blockchain can host data sharing, PKI services, and access controls, which have asymmetric requirements of reading and writing. This also explains why a considerable portion of blockchain-IoT papers in the survey~\cite{panarello2018blockchain} focus on these topics. 

Another approach is using private or permission blockchains instead of the public blockchain, where the burden of computing-intensive proof-of-work is relieved or avoided. However, it also undermines the security with a much smaller network scale. Some new blockchain techniques such as hyperledger~\cite{androulaki2018hyperledger} and IoTa~\cite{popov2016tangle} achieve better performance by making trade-offs either on decentralization or on security.

\subsection{Rethinking Blockchain and IoT Integration}\label{subsec:rethink}
\begin{figure}[t]
    \centering
	\includegraphics[width=0.45\textwidth]{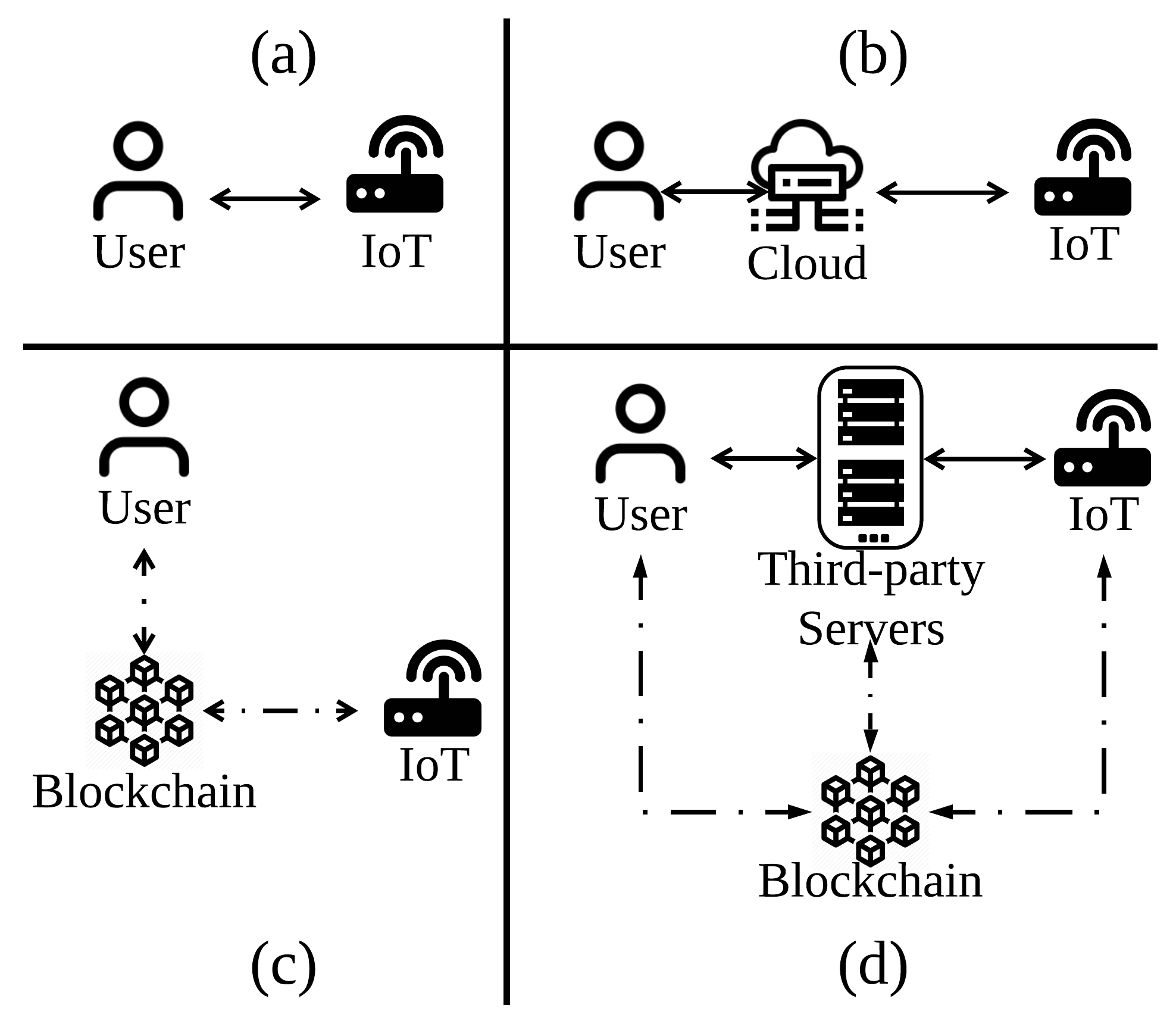}
	\caption{Abstract architectures of IoT services. Solid lines represent direct Internet communications and dash-dotted lines represent interactions with the blockchain.}
	\label{fig:abstractmodel}
\end{figure}

To explore a feasible way to integrate IoT and blockchain, we first make an abstract model of IoT services as depicted in Figure~\ref{fig:abstractmodel}. In a conventional IoT services scenario, with client devices (e.g., smartphone), users can either access their IoT devices directly (case (a)) or through the cloud server for advanced functionalities like cloud storage and automation (case (b)). The blockchain, as mentioned earlier enables IoT services to replace the cloud server with the blockchain (case (c)). One promising architecture is proposed in~\cite{zyskind2015enigma} as shown in (case (d)) that combines the ordinary distributed system with blockchain by moving heavy-load services from the blockchain to third-party servers. The blockchain serves as the coordinator to enforce correctness and fairness. However, this paper mainly focuses on the theoretical model of secure multi-party computing and does not offer more details of how to realize it in the IoT domain.



We follow this insight and dive into more details of blockchain-assisted distributed IoT services to find the answer to the three questions as proposed in the Section~\ref{sec:into}. Many blockchains provide the functionality of cryptocurrencies. So, the blockchain can be used as a service trading platform where service users (usually IoT devices) use services provided by third-party service providers who join for service fees paid by cryptocurrencies. To avoid possible disputations, the blockchain can also serve as the intermediary between two parties that collects pre-paid service fee payments from service users and compensates providers when objective proofs are provided. The blockchain can also punish malicious service providers by forfeiting their deposits of cryptocurrency to deter possible attacking attempts.

%% file: sections/casestudy.tex
\section{Case Study of Efficient Blockchain and IoT Integration: IoT Remote Access}\label{sec:case}
Smart home IoT systems enable homeowners to manage their IoT sensors and appliances from both inside and outside their homes. However, accessing IoT outside the home's private network is challenging due to the isolation of Network Address Translation (NAT) and the firewall. As a result, IoT remote access requires a relay server with a public accessible IP address for bridged communication. In this section, we study the use case of IoT remote accessing to demonstrate the proposed framework. 



\subsection{State-of-art Solutions}
\paragraph{Port forwarding}
Direct access through the IP addresses may be obstructed by the NAT and firewalls. They either shield IoT devices in private networks or filter out inbound connections. The intuitive workaround is exposing the device's port on the public address by setting the port forward on the gateway device. However, as pointed out in~\cite{sivaraman2016smart}, the NAT has side-effects of isolating IoT devices and protecting them from attacking traffics on the Internet. Some consumer IoT devices such as Belkin's WeMo smart plug and Philips Hue smart light accept unauthenticated commands from the local network and require no credentials to simplify users' configuration. However, exposing IoT devices them could be dangerous because any host on the Internet gets the possibility to compromise them. For example, a surprising higher number of IoT devices are infected by the Mirai botnet malware because they are exposed to the public Internet with the UPnP IGD protocol~\cite{krebs2016makes}.

\paragraph{Cloud-based Endpoint}
\begin{figure}
	\centering
	\includegraphics[width = 0.45\textwidth]{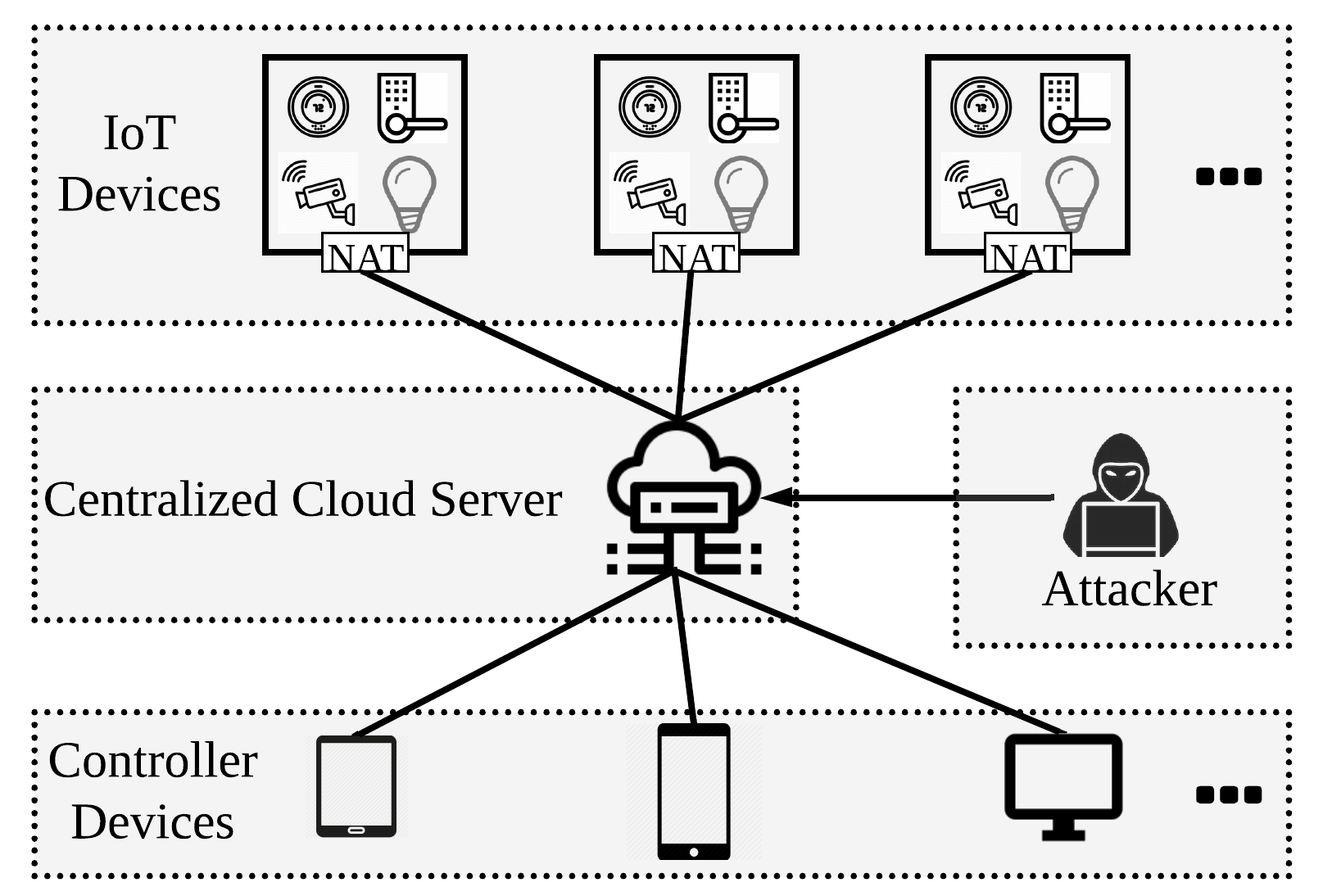}
	\caption{Relay-assisted IoT accessing model. For the scenario of both two parties are behind the NAT, a relay server with a publicly accessible IP address is necessary for bidirectional communications.}
	\label{fig:basic_relay}
\end{figure}
Even shielded by the gateway, vulnerable IoT devices behind the home private network are still threatened by other compromised devices through the LAN connection as described in~\cite{sivaraman2016smart}. Modern IoT vendors try to address this problem with a more radical isolation that anchors the access entry of their products on the endpoint instances hosted by cloud servers as shown in Figure~\ref{fig:basic_relay}. With the embedded public key, IoT devices can initiate a secure session towards the endpoint right after it boots up and maintain the session with heartbeat packets. This method enables indirect access through the relay server~\cite{ajitomi2015cost}. Since the session is proactively launched by the IoT device located behind the NAT, access requests encapsulated in the response packets can freely pass through the NAT and firewall, which is dubbed as "piggybacking." A typical real-world example is the message broker service~\cite{aws-broker} provided by Amazon Web Service's IoT core module. It maintains a long-term session with the subscribed IoT devices where messaging protocols such as MQTT are carried with it for real-time device control.

Counterintuitively, the centralized cloud could provide attackers, especially botnet attackers, better approaches to deliver IoT attacks. As the cloud server usually groups entries for similar devices from the same vendor, once it gets compromised~\cite{cloud-breach}, the attacker suddenly acquires free access to a large amount of IoT devices which share similar vulnerabilities. This saves attackers the effort for randomly scanning on the Internet, which is time-consuming. Even the cloud server is not taken down, security flaws of application program interfaces (APIs) as summarized in~\cite{alrawi2019sok} can also facilitate attacks like username enumeration, password brute-forcing, and unauthorized privilege escalation. These insecure cloud APIs prompt the cloud-based IoT botnet attacks~\cite{pierre2017dlink,kimp2p}. Aside from security issues, the scalability problem is another concern. Maintaining encrypted sessions involves the device identity management, the credential update and constant communication, which induce heavy pressure on the centralized cloud server. Considering the scale of IoT devices, centralized cloud servers can potentially become the performance bottleneck whose malfunction may result in a large-scale blackout of IoT devices.

\subsection{Threat Model}
According to the background knowledge and the aforementioned challenges, we present our threat model here. Following the \emph{access-to-attack} assumption, we assume very strong attackers who have capabilities to: 

\begin{enumerate}
    \item Take down any IoT devices  as long as they can get access to them by exploiting application layer software vulnerabilities.
    \item Take down arbitrary relay servers with non-negligible time.
    \item Install malwares on compromised IoT devices for botnet attacks.
\end{enumerate}



These capabilities are reasonable considering the endless discoveries of new vulnerabilities on vendor customized applications.  Also, we make some reasonable assumptions about the restriction of attackers' capabilities: 
\begin{enumerate}
    \item Attackers cannot take down an unknown device without making multiple attempts.
    \item Attackers cannot crack standard cryptographic primitives such as the digital signature.
\end{enumerate}

These assumptions are practical because of the principle of multi-version programming. IoT applications made by different vendors may have different vulnerabilities, but there is no universally applicable vulnerability considering the variety of IoT software. Besides, since the hardware, the operating system, and public middleware libraries such as TLS-N are widely shared by different device vendors and are well tested in contrast to individually developed application layer programs, it is much more difficult to find vulnerabilities to exploit among them. So, we think those extremely strong attacking vectors targeting IoT hardware and fundamental software components are out of our scope.



%% file: sections/mydesign.tex
\section{Design of RS-IoT}\label{sec:design}
In this section, we present our design of the blockchain based relay-sharing IoT remote access system (RS-IoT), which utilizes the framework we propose in Section~\ref{sec:integration}. Since a relay server is necessary for accessing IoT devices behind the NAT, a feasible solution is to decouple the fixed relationship between relay server and the IoT devices by replacing the centralized cloud server with a large number of third-party relay servers. Accordingly, a smart contract based service trading platform is designed for transaction management and dispute arbitration, with which IoT devices are able to freely choose their relay service providers. 

Our design brings four prominent benefits: 1) The management platform is completely self-governing to guarantee the fairness of the service trading without the help of any trusted authority. 2) Compromised or misbehaving relay servers will be reported and excluded from the relay platform to avoid further damage. 3) Malicious relay servers that launches attacks against its connected IoT devices have the risk to be reported and punished 4) No assumption of robust and secure IoT applications are required. 




\begin{figure}[ht]
    \centering
    \includegraphics[width = 0.45\textwidth]{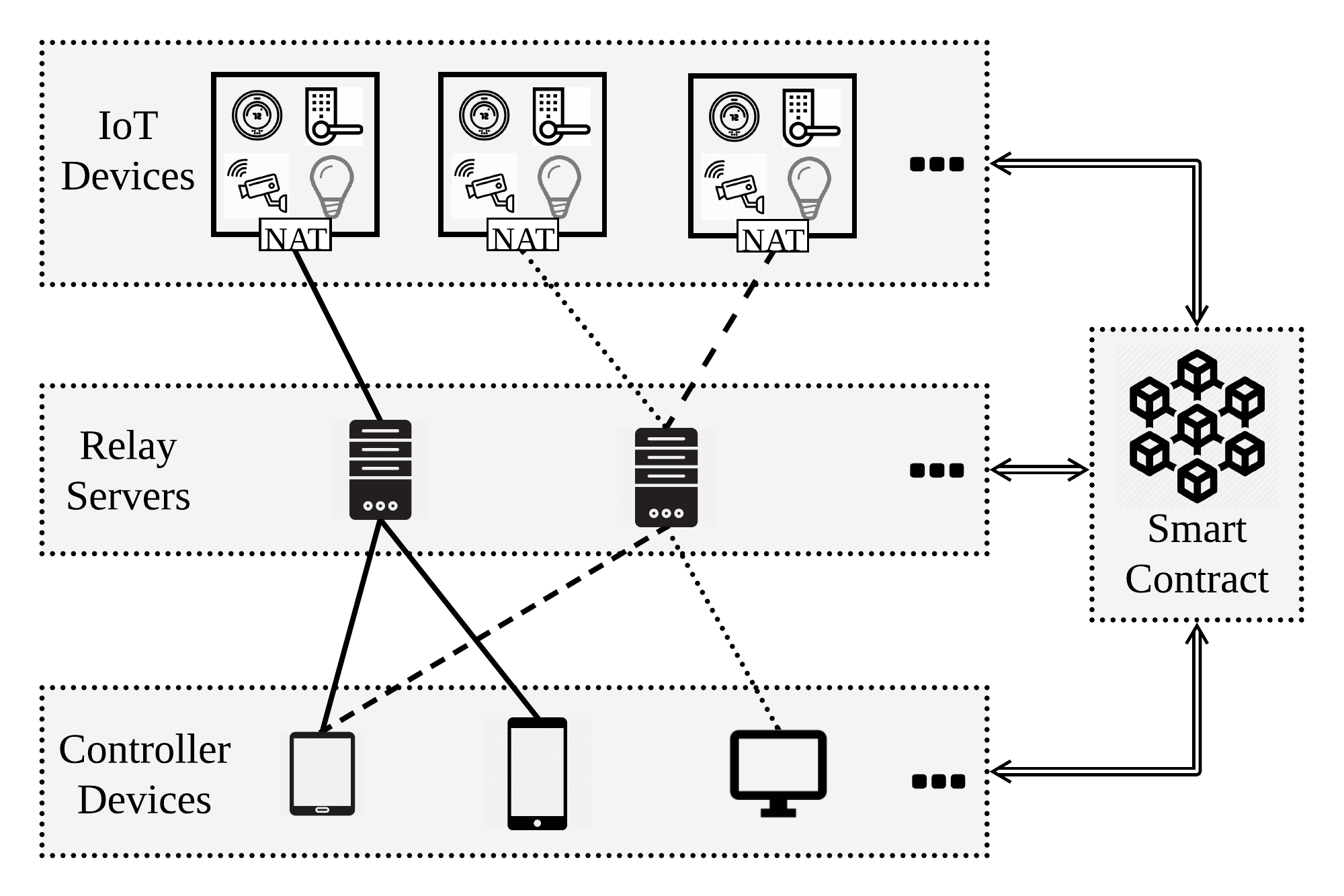}
    \caption{Overview of relay-sharing IoT remote accessing. Nodes connected by the same type of line are within the same session of a relay service. A relay server can host multiple control sessions and a controller device can control multiple IoT devices through different relay servers.}
    \label{fig:design}
\end{figure}

\subsection{Relay Sharing Model}\label{sec:relay_sharing_model}
Taking the smart contract into account, there are four roles involved in RS-IoT as shown in Figure~\ref{fig:design}: the IoT device ($D$), the controller ($C$), the relay server ($R$), and the smart contract ($SC$). Among them, the controller and the IoT device are grouped as the party of the service user and share secret keys since they are usually owned by the same owner and have the common interest. The relay server constitutes the party of service provider. The smart contract is script code stored on the public blockchain as a fair third-party.

All roles except the smart contract join the blockchain by generating their own public and private keys, and the smart contact is published on the blockchain with only the account address. We denote their addresses on as $addr(D)$, $addr(C)$, and $addr(R)$ respectively. After that, the IoT device, the controller and the relay server top up their account with cryptocurrency as deposit or service fee. The basic unit of relay service is the end-to-end link between an IoT device and its controller client with one commissioned relay server in between. Both the IoT device and the controller are connected to the relay server via TLS-N sessions and all packets passed through them are signed with the sender's blockchain private key. 



\subsection{Relay Workflow}
\begin{figure*}
	\centering
	\includegraphics[width = 0.9\textwidth]{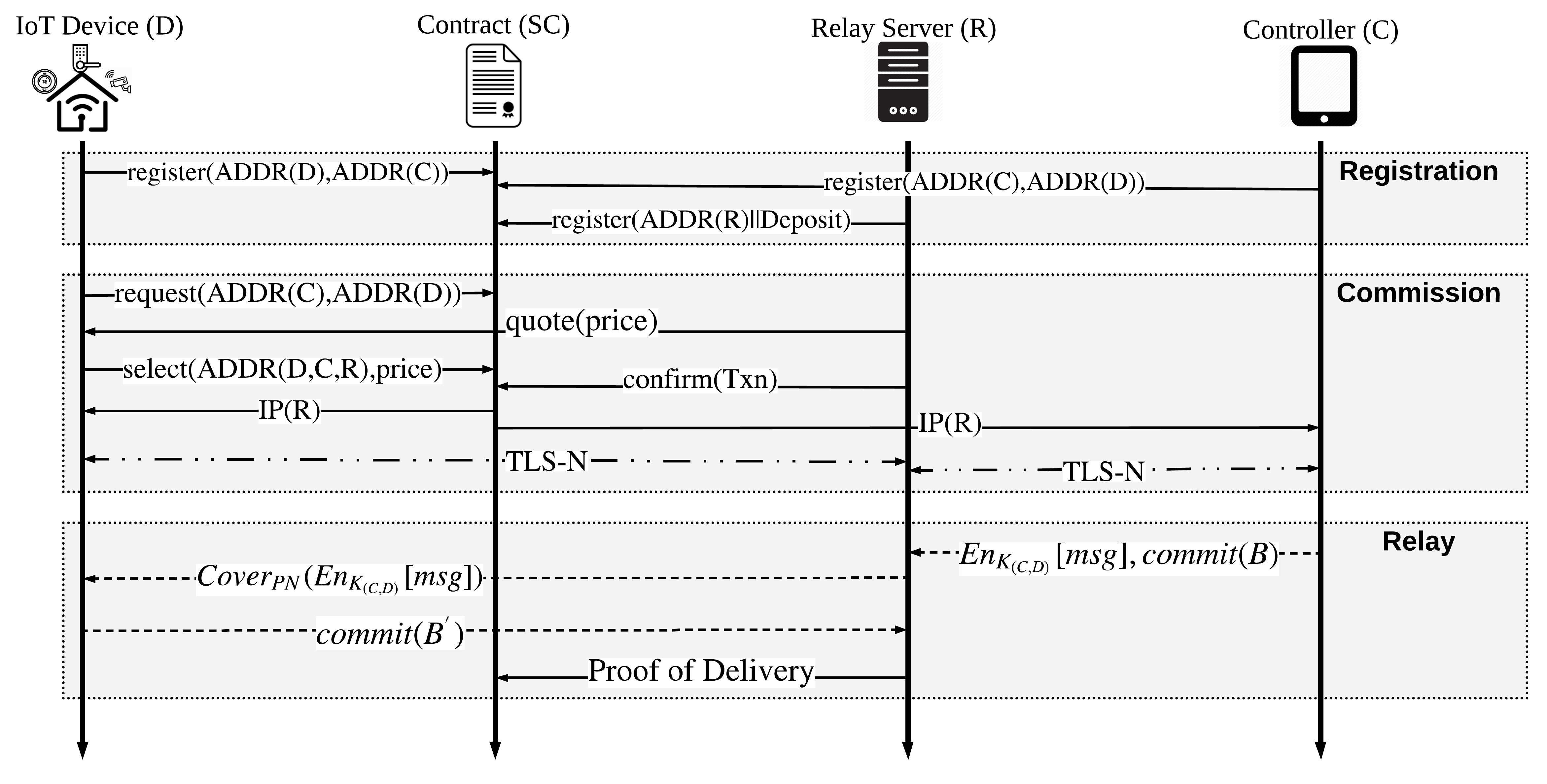}
	\caption{Workflow of RS-IoT. Solid lines are function call transactions toward the smart contract and blockchain direct transactions. Dash lines are direct communications through the TLS-N session.}
	\label{fig:workflow}
\end{figure*}

This subsection describes the workflow of relay sharing by going over the procedure of forwarding a packet. As described in Figure~\ref{fig:workflow}, the workflow consists of the registration phase, the commission phase, and the relay phase.

\subsubsection{Registration}
With blockchain accounts established, all three parties register their account addresses on the smart contract to indicate their participation by calling a registration function of the smart contract $SC$ and creating new items in two tables in blockchain: \textit{userInfo} and \textit{serverInfo}. Since a service user is uniquely identified by its pair of IoT device ID and controller ID, both of them need to register by providing the other party's address as arguments. After either one of them calls the registration function, a new item in \textit{userInfo} is created with the confirmation flag set to false. Then, the function call by the other party would flip the confirmation flag to indicate a successful user registration. For the relay server, the deposit of cryptocurrency is required to be paid to the smart contract along with the registration function call transaction. The amount of deposit is stored in \textit{serverInfo} together with the relay server's blockchain address.

The function prototype for registration is listed below
\begin{lstlisting}[language=Solidity]
function reg_user(address oppo_end) public {...}    
function reg_server() public payable{...}
\end{lstlisting}

\subsubsection{Commission}
The commission phase is used for mutual discovery between service users and providers, as well as setting up service relationships. The commission phase begins with an IoT device calling the \textbf{service request} function which broadcasts a global event containing the user's registration information. Upon receiving the event, interested relay servers respond with their IP addresses and quotes of service via direct transactions towards the requesting IoT's blockchain address. After waiting for some time, the IoT device evaluates the received quotes by the deposit and price, and finally makes a decision by calling the \textbf{service select} function with the chosen relay server's blockchain address and price as arguments. Similar to the user registration, this function call would generate an item into the table \textit{serviceList} consisting the following values as shown in Table~\ref{tab:keys} and set the confirmation flag to pending.

\begin{table}[hbp]
	\caption{Keys of content in the serviceList.}
	\centering
	\scalebox{1}{
		\begin{tabular}{l l p{10cm}}
			\toprule
			\centering
			\textbf{$Txn$} & index number for each pair of IoT device and relay server\\
			\textbf{$Serial$} & counter to index the number of successfully relayed packets\\
			\textbf{$Address$} &  blockchain address of all involved parties\\
			\textbf{$Price$} & cost of forwarding one packet\\
			\textbf{$Balance$} & amount of pre-paid service fee payment\\
			\bottomrule
		\end{tabular}
	}
	\label{tab:keys}
\end{table}


Finally, the chosen relay server confirms the service relationship by calling the \textbf{service confirm} function to change the confirmation flag in table \textit{serviceList} to confirmed. At the same time, an event broadcast would be triggered as a log of relationship binding. Then, the commission is finished. The IoT device launches a TLS-N connection towards the commissioned relay server.

The function prototype for commission is listed below
\begin{lstlisting}[language=Solidity]
function service_request(address D, address C) public {...}    
function service_select(address D, address C, address R, uint price) public payable{...}
function service_confirm(uint Txn){...}
\end{lstlisting}

\subsubsection{Relay}\label{subsec:relay}
As all history transactions on blockchain are publicly readable, the controller client can easily recover the current serving relay server's IP address and initiate a TLS-N connection towards it. With the shared secret key between the IoT device and the controller client, a long-term symmetric encryption key $K(C,D)$ can be derived to encrypt the messages exchanged between them. The controller client first generates a proof of signed transaction on the original packet which is sent to the relay server along with the packet itself. On receiving the packet, the relay server uses a random stream derived from a one-time key $PN$ to cover the packet before forwarding it to the IoT device. Afterwards, the receiver (IoT device) generates another proof transaction using the same algorithm but on the covered packet and sends it back to the relay server. Finally, the relay server creates a new transaction containing the cover key and  broadcasts it along with two received proof transactions. With the cover key, a successful proof verification on the smart contract will trigger a transfer of cryptocurrency from the contract account to the relay server's personal account. We will illustrate details of the proof generation algorithm and show its effect on preventing cheating in Section~\ref{sec:pod} and Section~\ref{sec:secanalysis}, respectively.

\subsubsection{Decommission}
As we stated, both the relay server and the IoT device can freely determine when to end the service relationship. The decommission process is provided to terminate a service relationship by either party by calling the function \textbf{decommission}. $Txn$ is the only argument required to specify the service record to be cleared. After a decommission is initiated, an event would be emitted as the notification and the service record item in \textit{serviceList} is deleted after a pre-defined block time for the relay server to finish billing. The remaining pre-paid service fee will be paid back to the IoT device's personal account.

\subsection{Relay Service Client}\label{sec:middleware}
\begin{figure}[ht]
	\centering
	\includegraphics[width = 0.4\textwidth]{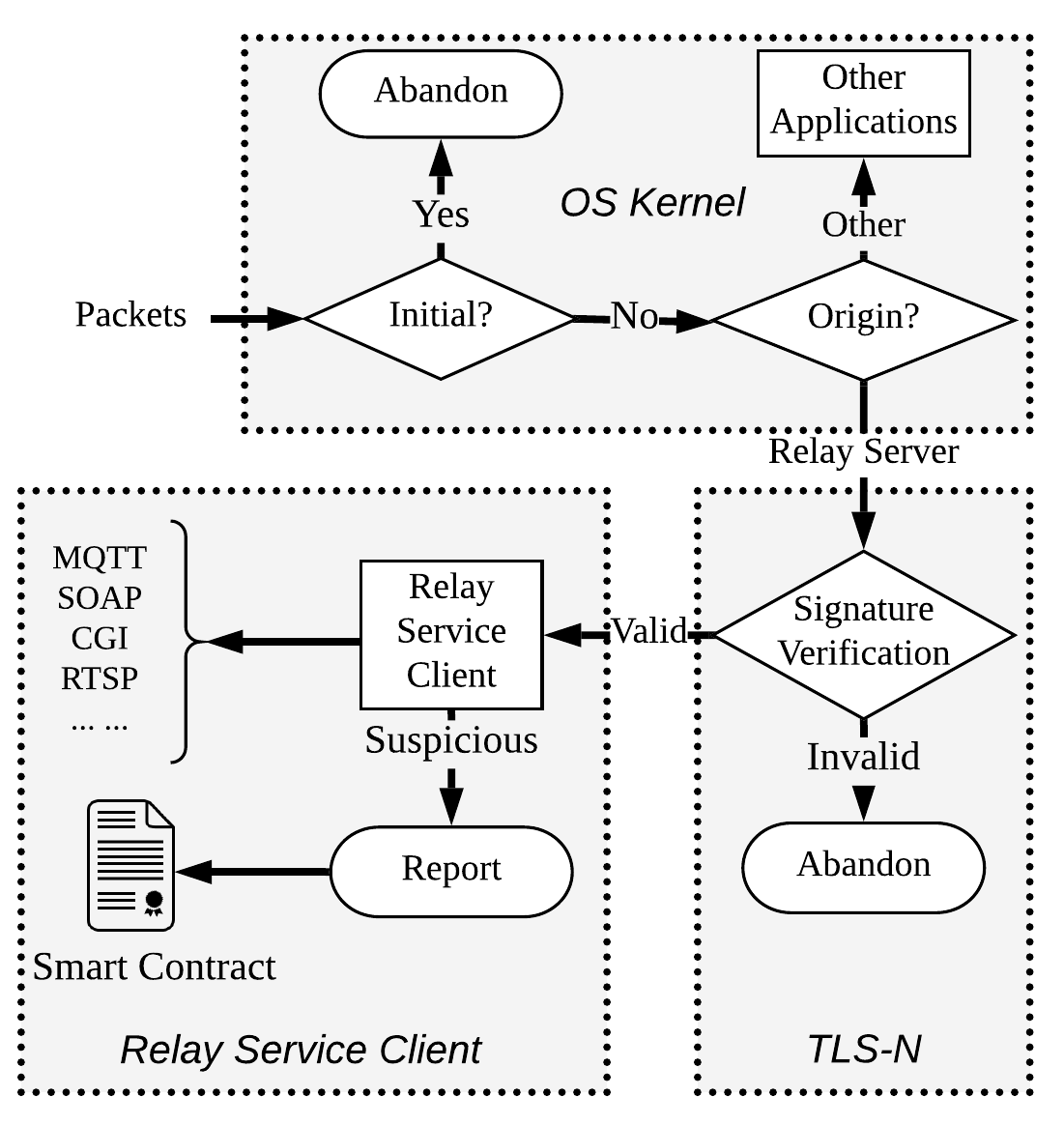}
	\caption{Packets handling procedures of IoT devices. The operating system can filer out traffic that does not originate from the commissioned relay server and the TLS-N library blocks packets without valid signatrues.}
	\label{fig:packet_handler}
\end{figure}

Relay Service Client is a middleware installed in IoT devices to deal with all relay sharing related logics. Besides interacting with the smart contract for the aforementioned workflow, this middleware is also responsible for setting the packet filter policy for the operating system kernel and configuring the TLS-N module with a private key. As shown in Figure \ref{fig:packet_handler}, for any incoming packets, the netfilter of the operating system first filters out those connection requests initiated by external hosts. Then, the netfilter dispatches packets that originate from the commissioned relay server to the TLS-N library which checks their signature. Only packets with valid signatures can be accepted and passed to the relay service client. Here, the client inspects received packets and report those suspicious ones to the smart contract. Different device vendors have different implementations of the relay service client and inspection method. 


\subsection{Billing of Relay Service}\label{sec:billing}
During the commission precedure, the pre-paid service fee is transferred to the smart contract's account along with the function call of \textbf{service confirm}. As described in Section~\ref{subsec:relay}, a relay server gets remuneration by presenting the proof of each successfully relayed packet.  However, considering the amount of relayed packets and blockchain's aforementioned cost and throughput constraints, verifying each of them on the smart contract is unrealistic. We innovatively use the smart contract's local verifiability to make it unnecessary to verify each packet on the blockchain. That is, on receiving proof transactions, the relay server first runs the verification function locally instead of posting them on the blockchain. If the verification is successful, the relay server caches the proofs and for the current packet, and send the cover key to the IoT device. If both two parties are honest, the IoT device would be able to successfully recover the message with the received cover key. As long as no dispute occurs, this offline verification can be used for all following packets. When the relay server wants to cash the remuneration, it only needs to verify the proof of the last packet on the blockchain. The difference between the recorded serial number $serial$ in the table \textit{serviceList}  and $serial$ in the proof is accounted as the number of successfully relayed packets. The total amount of remunerations is then calculated as the product of the number of successfully relayed packets and the unit service price. After the transfer, the $serial$ in \textit{serviceList} is updated. Because the $serial$ as a function call argument is signed by the IoT device, it can be regarded as the IoT's acknowledgment of successful relay for all packets before it.

%% file: sections/proofofdelivery.tex
\section{Proof-of-delivery}\label{sec:pod}
In a real-world service trading system, both the service user and the service provider have the incentive to cheat: the relay server may deliver broken, modified, or forged packets for making extra profit; While, the relay user (including the controller and the IoT device) may deny the receipt of packets to avoid the payment. 

To solve this problem, we propose an autonomous proof-of-delivery solution to resolve possible disputes fairly by utilizing the smart contract as a decentralized trusted computing platform. Firstly, We design a SHA-3~\cite{dworkin2015sha} based key stream generator for the relay server to hide the content of the original packet. Then, leveraging the smart contract's locally verifiable feature, we propose an innovative off-line blind proof generation algorithm to derive proofs of packet delivery on both the original and the covered packet. During the operation, the relay server holds the cover key while it asks the IoT device for the proof of the covered packet. On one hand, without receiving the correct proof, the relay server would not reveal the cover key for the receiver to extract the message. On the other hand, the relay server is not able to get the correct proof for cashing reward if the packet is not delivered confidentially. This solution provides a mutual restriction between the user and the server so that neither of them have the opportunity to cheat. Thanks to the smart contract's off-chain verifiability, the proof only needs to be posted on the blockchain when disputes occur rather than every time a packet is relayed, which greatly reduces the cost of  operation. For clarity, we list all symbols to be used in the following notation table:

\begin{table}[htbp]
	\caption{Notations}
	\centering
	\scalebox{1}{
		\begin{tabular}{r c p{10cm}}
			\toprule
			\centering
			$K_{(C,D)}$ & pre-shared encryption key between $C$ and $D$\\
			$S_{(C,D)}$ & pre-shared secret for bits selector\\
			$Ra\ \&\ Ra'$ &  byte select index list\\
			$PN$ & secret key for cover stream generator\\
			$B$ & selected bytes for uncovered packet\\
			$B'$ & selected bytes for covered packet\\
			$Tx(B)$ & contract function call with $B$ as argument\\
			$N$ & commitment length\\
			$i$ & self-incrementing counter\\
			\bottomrule
		\end{tabular}
	}
	\label{tab:TableOfNotationForMyResearch}
\end{table}

\begin{figure}[ht]
	\centering
	\includegraphics[width = 0.45\textwidth]{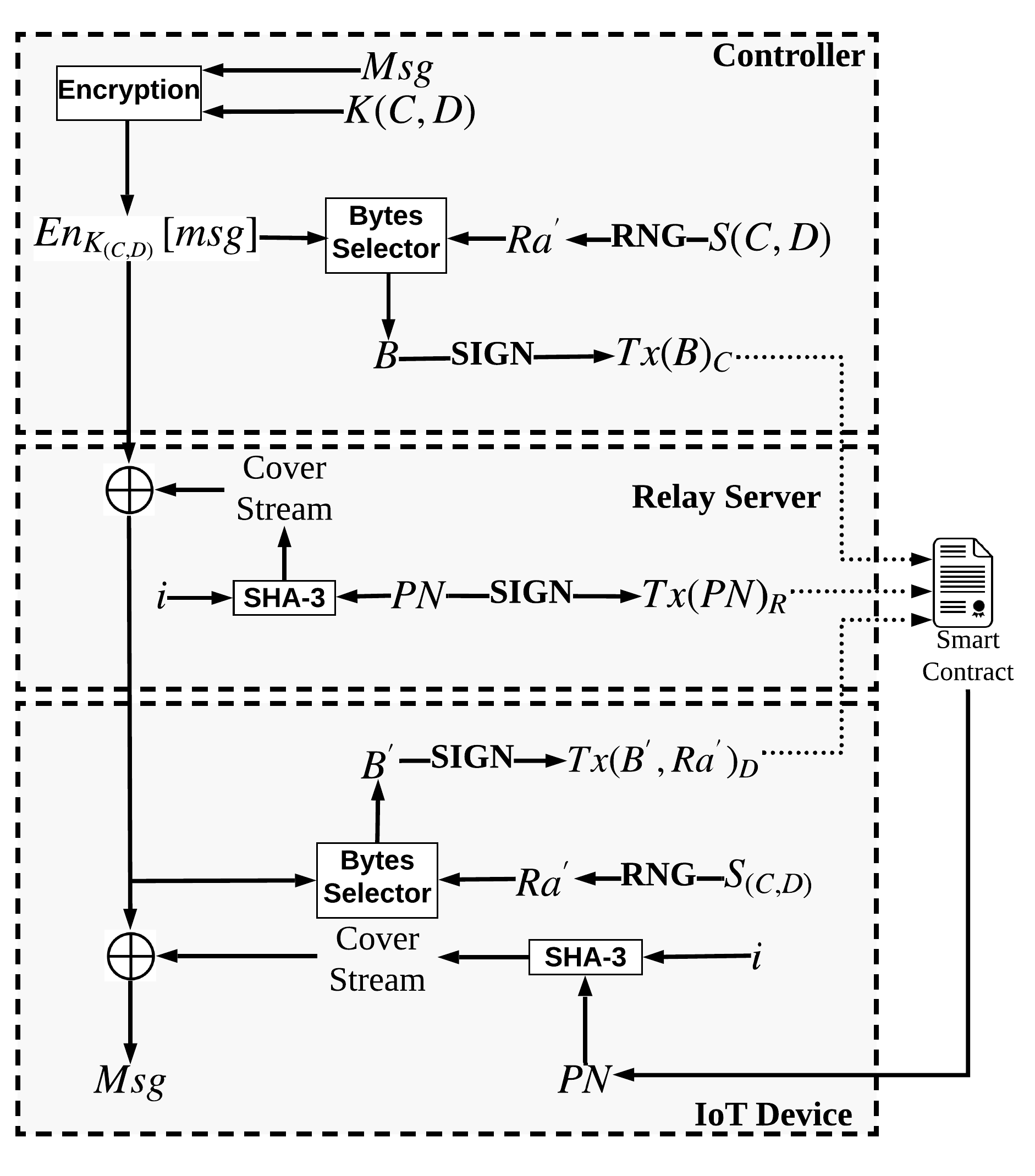}
	\caption{The workflow of proof-of-delivery. Dotted lines mean signed transactions are sent to the relay server. After the verification, the relay server publishes them on the smart contract to get its service fee.}
	\label{fig:pod}
\end{figure}

\subsection{Components}
Considering the high cost of storing and processing data on the blockchain, it is not practical to verify the entire packet on smart contract because of the unbearable latency and cost. Although existing cryptography primitives have already provided satisfying digest-based security features, hardly any of them are available as built-in functions on popular public blockchain platforms such as the Ethereum~\cite{wood2014ethereum}. If we implement them in the form of script code, the cost will become unbearable.  To overcome these difficulties, we design the \textbf{Bytes Selector} and the \textbf{Cover Stream Generator} as new primitives based on Ethereum's built-in functions. The Bytes Selector is used to extract fixed-length bytes streams from arbitrary packets, while the Cover Stream Generator is used to generate cover streams to hide the content of a packet. These new components provide comparable security while remains low cost.

\subsubsection{Cover Stream Generator}
To prevent the lost of service fee when service users maliciously deny the receiving of relayed packets, the relay server covers the content of packets to be delivered to the receiver with a stream cipher. The relay server will only reveal the cover key when it gets valid commitments from the receiver as the proof of successful delivery. The cover stream generator is an alternative implementation of the stream ciphers on the smart contract. Since there is no pre-compiled script function in the current version of Ethereum blockchain, implementing a standard stream cipher would be extremely expensive. Under this case, we build the cover stream generator based on the SHA-3 hash function, which is the cheapest built-in operation on Ethereum~\cite{wood2014ethereum}. 

The cover stream is generated by concatenating hashes of the sums of a secret key $PN$ and a self-increasing counter $i$. To keep high entropy of randomness, we only retain the highest indexed byte of each 256-bit SHA-3 hash result. The cover stream is generated as shown in the equation below. The `$|$' means the concatenation of hash values.
\begin{equation}
cover(PN) = SHA3(PN)|SHA3(PN+1)|\cdots
\end{equation}

Aside from using the low-cost building block, this hash chain based stream cipher also reduces cost by enabling selective encryption stream generation at arbitrary position. For example, to encrypt/decrypt the content at the $K$th byte with a given key $PN$, we can simply calculate $SHA3(PN+K-1)$ and take the first bytes as the key stream instead of producing the key stream from the beginning. Thus, the overhead from generating and storing the whole stream for large packets is avoided.

\subsubsection{Bytes Selector}
As an analogy, the bytes selector serves the similar purpose as the Hash Message Authentication Code (HMAC) function. The difference is that our bytes selector retains the commutativity along with our cover scheme (i.e., the digest of the covered packet equals to the covered digest giving same secure keys).

The bytes selector is driven by the secret $S(C,D)$ shared between the controller client $C$ and the IoT device $D$. Both $C$ and $D$ use this secret as the seed of a pseudo random number generator. For each packet, $C$ and $D$ synchronously generate $N$ random numbers of 16-bits, denoted as $Ra = \{ra_1,ra_2,ra_3,\cdots,ra_N\}$. Assuming the length of the packet is $L$, the bytes selection list $Ra' = \{ra'_1,ra'_2,ra'_3,\cdots,ra'_N\}$ is derived from $Ra$ with $ra'_i = ra_i \mod L$. Thereafter, a list of $N$ bytes $B=\{b_1,b_2,b_3,\cdots,b_N\}$ are extracted from targeting packet where $b_i$ is the $ra'_i$th byte in the packet. The generated bytes list is totally random and there is no way to recover their locations in the packet without the knowledge of the random seed $S(C,D)$.



\subsection{Proof-of-Delivery Workflow}
The proof-of-delivery comprises four steps: 1) The  sender (assuming it is the controller because control session is usually initiated by it) generate the first commitment with our bytes selector on the original packet to be sent to the relay server. 2) The relay server covers the packet and forwards it to the receiver (IoT device). 3) The receiver generate the second commitment on the covered packet using the same methond and sends it back to the relay server. 4) The relay server verify two received commitments with the cover stream key and reveal the key to the receiver if commitments are valid.

\subsubsection{Bytes Commitment by the Controller Client}
We assume the controller $C$ wants to send a packet $msg$ to the device $D$ through the relay server $R$. $C$ firstly encrypts the message with the pre-shared symmetric key $K_{(C,D)}$ and generates the encrypted packet $En_{K(C,D)}[msg]$. Then, it rolls the random number generator with the pre-shared secret $S(C,D)$ to get the $Ra'$ as indices of bytes to be selected. After that, it extracts the bytes from  $En_{K(C,D)}[msg]$ as indexed by $Ra'$ to form the commitment $B$ and use it as the argument of the transaction towards the function of \textbf{commitment}. Finally, the transaction $Tx(B)_C$ is signed by the controller client's private key and sent to the relay server $R$.



\subsubsection{Bits Covering by the Relay Server}
On receiving the encrypted packet $En_{K(C,D)}[msg]$ and the commitment $B$ from the controller client $C$, the relay server $R$ generates a random number $PN$ as the seed of the cover stream generator to produce an pseudo random stream. It uses the cover stream to cover the packet with the exclusive-OR operation. which is illustrated as in the equation below. After that, the covered packet is forwarded to the IoT device. 
$$Co(En_{K(C,D)}[msg]) = En_{K(C,D)}[msg]\ \oplus\ cover(PN)$$



\subsubsection{Bits Commitment by the IoT Device}
The received packet from the relay server is covered by the cover stream. By using the bytes selector, bytes at the same location are selected on the covered packet which is denoted as $B'$. Then, the IoT device packages $B'$ together with the bytes selection list $Ra'$ into a function call transaction $Tx(B',Ra')_D$. The signed transaction is sent back to the relay server $R$.


\subsubsection{Asynchronous Delivery Verification}\label{sec:deliver}
Upon receiving both commitment transactions $Tx(B)_C$ and $Tx(B',Ra')_D$, the relay server now has all the of materials to verify the correctness of the commitments locally. Then, the relay server checks whether $B\ \oplus\ B' $ equals to $cover(PN)$ at the designated location as specified by $Ra'$. If the verification is successful, the relay server prepares another transaction $Tx(PN)_R$ signed by itself with the $PN$ as the argument. These three commitment transactions are the proof of delivery which are cached by the relay server. When commitments are presented to the smart contract $SC$, the same verification is performed as the relay server. Payment for the relay service is transferred to the relay server upon successful commitment verifications.

To avoid the latency and the transaction fee caused by executing the smart contract, the relay server caches all commitments instead of verifying them immediately. It delivers the cover key $PN$ through the relay connection to the IoT device after the successful verification. When it wants to withdraw the payment, it verifies the commitment of the newest packet. As described in Section~\ref{sec:billing}, the verification of all previous commitments is unnecessary.



%% file: sections/penalty.tex
\section{Penalty \& Disputation Solving}\label{sec:penalty}
Since the relay server serves as the only access entry for all its customer IoT devices, all traffic delivered to the IoT device should originate from its legitimate controller clients. However, relay service providers are anonymous according to the registration phase, which induces the risk of malicious relay servers delivering IoT malware. Based on the non-repudiable features of TLS-N, we develop a smart contract function for relay service users reporting unauthorized traffic from their commissioned relay server. After issuing a log event as notification, the reported relay server needs to present the TLS-N signature of the sender for the reported packet to claim its innocence. Otherwise, the relay server's registration will be revoked, and all its deposit will be confiscated.



\subsection{Reporting}[ht]
\begin{figure}
	\centering
	\includegraphics[width = 0.45\textwidth]{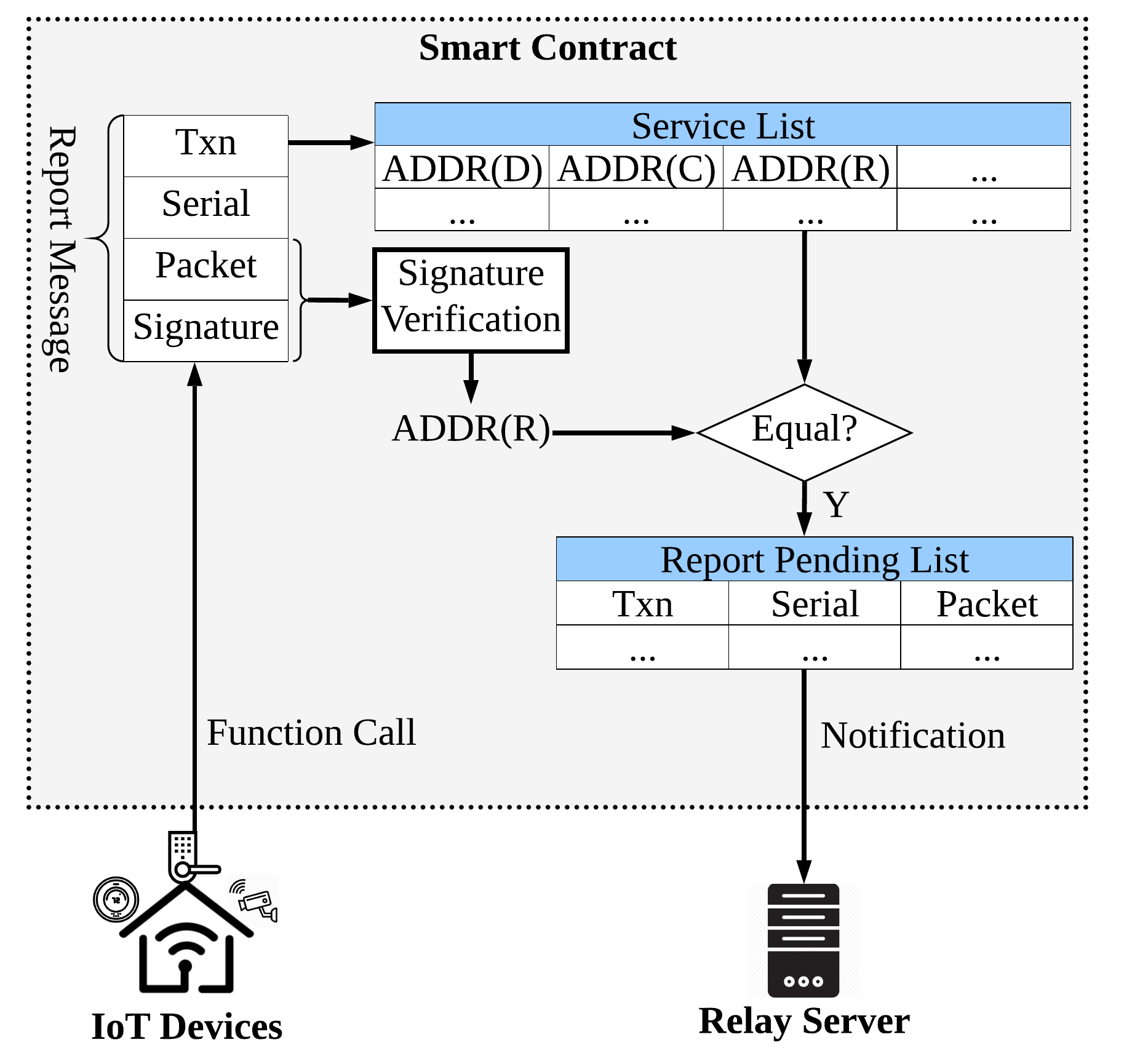}
	\caption{The workflow of reporting suspicious packets. Once the public key derived from the signature matches the address of the accused relay server, an record is inserted into the pending list. At the same time, it triggers an event to notify the relay server.}
	\label{fig:report}
\end{figure}
Upon receiving a packet, the IoT device firstly verifies the TLS-N signature with the TLS-N library to make sure it really originates from the relay server. Then, the TLS-N library passes the packet to the application layer of the IoT devices. If the packet contains malicious content that is not initiated by the controller, IoT devices have some chance to recognize it. According to the theory of multi-version programming (MVP), a perfect attack that can compromise arbitrary IoT devices does not exist considering the variety of IoT software types and architectures. Once a malicious packet fails to take down a device, the device can use the relay server's TLS-N signature to report this misbehavior to the smart contract.

The reporting process starts from the function call transaction towards the \textbf{reporting} function in the smart contract. As shown in the figure~\ref{fig:report}, the function call contains arguments of the transaction number, the suspected packet's serial number, the content of the packet and the relay server's TLS-N signature. Since the TLS-N signature is signed with the relay server's private key, when giving the content of the original packet, any arbitrary party including the smart contract can retrieve the relay server's public key and then derive the blockchain address $ADDR(R)$. The smart contract will verify the validation of the signature by generating the address from it and compare it with the address stored inside the \textit{serviceList} which is indexed by the argument of the transaction number. If the recovered address matches that in the list, it means the reported packet really comes from the relay server and the smart contract will take it as a valid report. Then the smart contract will create a new record in the report pending list which contains the transaction number, the serial number, the reported packet, and the current latest block number. At the same time, a notification is emitted and broadcasted on the blockchain to inform the relay server to process this accusation.


The function prototype for reporting is listed below:
\begin{lstlisting}[language=Solidity]
function reporting(uint Txn, uint serial, bytes packet, bytes32 signature){...}
\end{lstlisting}

\subsection{Rebutting}[ht]
\begin{figure}
	\centering
	\includegraphics[width = 0.45\textwidth]{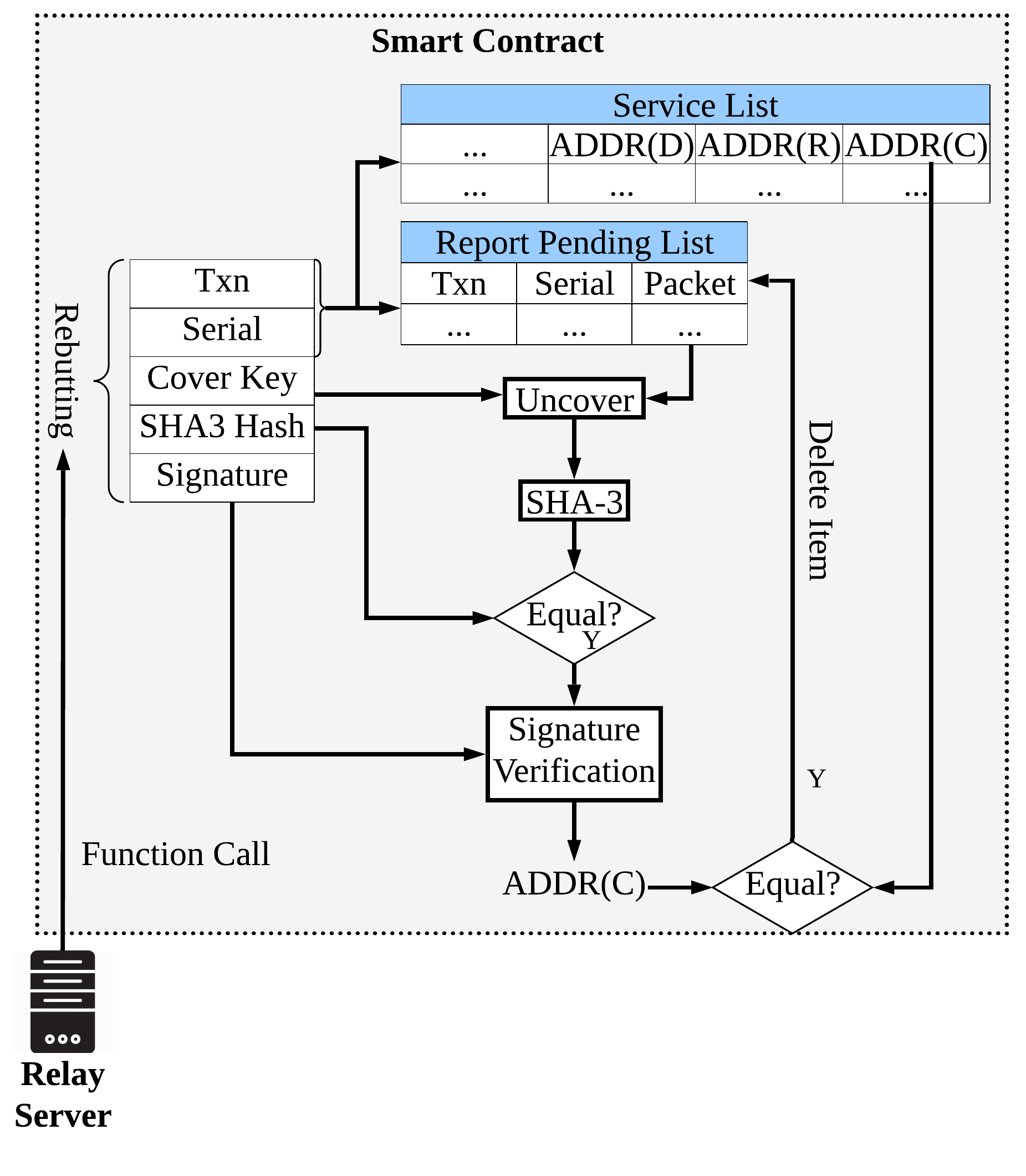}
	\caption{The workflow of rebuting a pending report record. The relay server prove itself's innocence by presenting the sender's TLS-N signature of the suspected packet and the cover key so that the smart contract can uncovered the packet and verify the TLS-N signature.}
	\label{fig:rebut}
\end{figure}
Even if the pending report record is successfully generated, it only represents the reported packet is sent by the relay server and is not enough to determine whether it is malicious. As a result, we design the \textit{rebutting} function for the accused relay server to defend its innocence by proving that the reported packet is originating from the controller device. 

As shown in the Figure~\ref{fig:rebut}, it involves a similar process that starts from sending the function call transaction towards the rebutting function in the smart contract. Among parameters provided, the transaction and serial number are used to locate the pending record of rebutting. Since the packet in the pending list is covered by the relay server with the cover key, the relay server needs to provide the cover key to recover the original packet it receives from the controller. Afterward, with the original packet and the controller's signature, the controller's address $ADDR(C)$ can be derived using the same way as in the reporting process. If the derived address matches that stored in the \textit{serviceList} table, it indicates the reported packet is really initiated by the controller, and the relay server does not do anything malicious. If so, the smart contract will delete the record in the report pending list. Otherwise the rebutting fails and the record remains inside the pending list.

The function prototype for rebutting is listed below:
\begin{lstlisting}[language=Solidity]
function rebutting(uint Txn, uint serial, bytes32 packetHash, uint PN){...}
\end{lstlisting}

\subsection{Executing}
Since the reporting notification takes some time to broadcast on the blockchain, a grace period is provided for the relay server to respond to the accusation.  The grace period can be measured with the number of newly generated blocks in the blockchain because the generation of new blocks usually means the consensus among all participating nodes in the network and the completion of states update. 

If the reporting record remains in the pending list after the end of the grace period, the IoT device who initiate the reporting is eligible to execute the penalty by sending function call transactions towards the \textit{executing} function in the smart contract. This time only the transaction number and the serial number is needed to locate the record in the pending list. The smart contract firstly traverses the report pending list to check the existence of the referred reporting record. Then, it calculates the number of blocks that are newly generated after the reporting. If the difference is larger than the pre-defined grace period, the smart contract will execute the penalty that means the relay server's deposit stored inside the smart contract account will be transferred to the reporting IoT device's account. Moreover, the misbehaving relay server's registration information in the \textit{serverInfo} list will be deleted to revoke its qualification as a relay server. As a result, it won't be able to be commissioned by any other IoT device in the future. Since the public blockchain is anonymous, the malicious relay server may rejoin the relay system with a different address to continue its attacks. However, registering as a new relay server requires the attacker to pay the deposit again, which causes economic loss to the attacker.

The function prototype for executing is listed below:
\begin{lstlisting}[language=Solidity]
function execute(uint Txn, uint serial){...}
\end{lstlisting}


%% file: sections/analysis.tex
\section{Security Analysis}\label{sec:secanalysis}
\begin{figure}
	\centering
	\includegraphics[width = 0.45\textwidth]{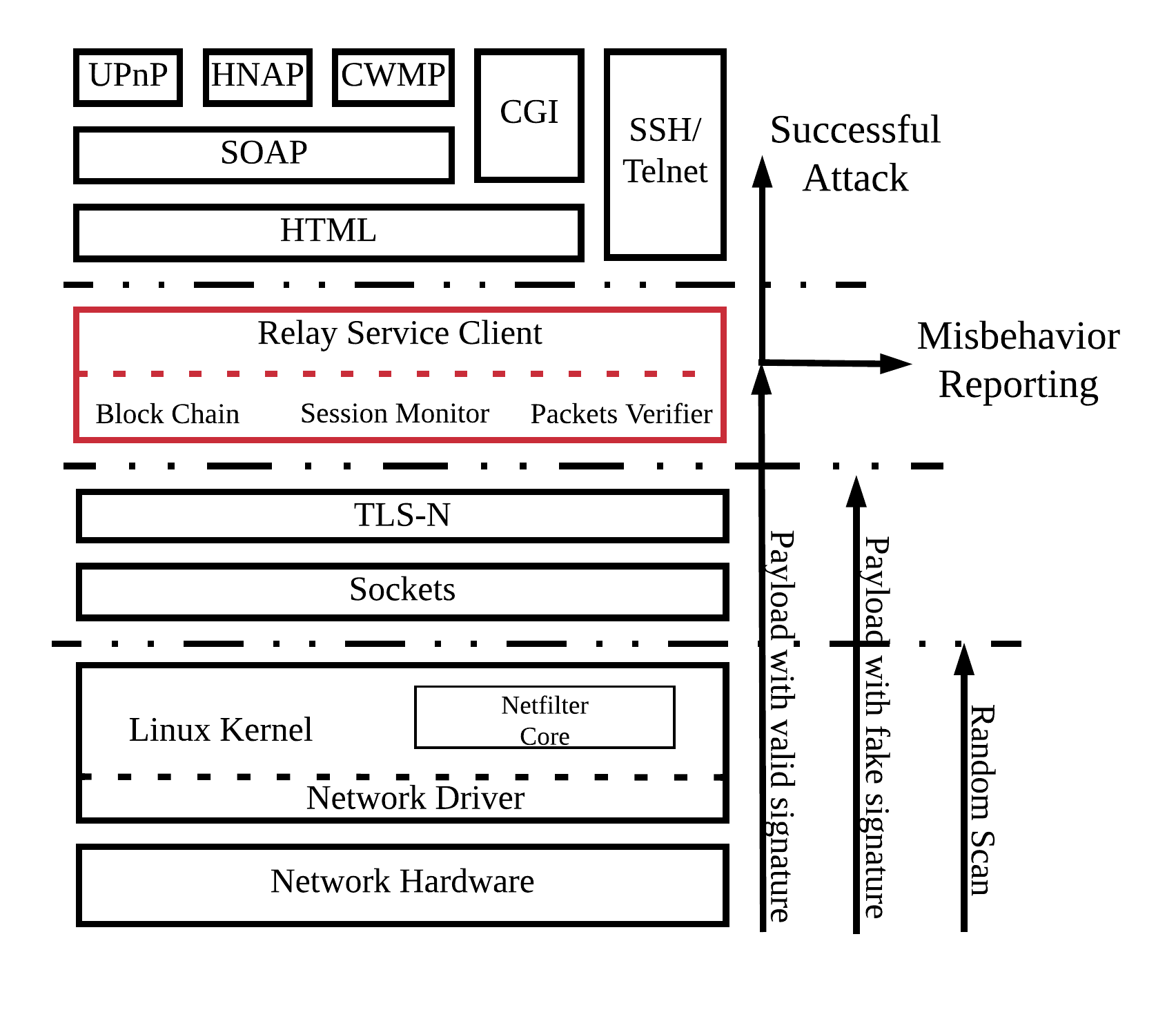}
	\caption{Demonstration of possible attacks.}
	\label{fig:attack_demo}
\end{figure}

\subsection{Possible Attacks}
\subsubsection{Random Scanning Attack}
In a random scanning attack, the attacker scans for open ports as other botnet malwares do without joining the relay sharing system. Since IoT devices are shielded by NAT gateways, scanning traffic from the Internet will not be able to reach the IoT devices. Even if the attacker successfully compromises devices located in the same subnet, the non TLS-N traffic will be discarded by the packet filter that resides in the IoT devices' operating system kernel. As depicted in Figure~\ref{fig:attack_demo}, this type of attack will never get a chance to exploit any vulnerabilities in the application layer.

\subsubsection{Attacks against the Relay Server}
In the centralized relay model, the relay server becomes an attractive target for attackers because once it gets compromised, attackers can efficiently access millions of its connected IoT devices. Differently, in our relay sharing system, attacking relay servers is much less efficient because each relay server only connects to a small number of IoT devices. In consequence, attackers need to spend significantly more efforts to take down many relay servers with different software implementations, while obtaining much less retribution. Even if some relay servers get compromised, they will be detected and excluded from the relay system when they are utilized for launching attacks.


\subsubsection{Malicious Relay Server Attack}
In this kind of attack, the malicious relay server attacks its connected IoT devices by delivering packets containing attacking vectors. According to our threat model, the malicious relay server must sign the packets with its blockchain private key to get it accepted by the target IoT device. As shown in Figure~\ref{fig:attack_demo}, the signed packet is then passed to our relay sharing client middleware.  The middleware may be vulnerable which means a malicious relay server does have the chance to successfully bypass the middleware's inspection and takes down the device without getting reported. However, as one of the core benefits of our work, we propose to deter this kind of attacks by imposing ``\emph{economic risk}'' instead of relying on the unrealistic perfect software implementation. Launching attacks inevitably requires the target's platform information which is acquired by sending some probing packets. If the target is not vulnerable to the probing attack, the unauthorized packets will be reported, which results in the malicious relay server losing all its deposit. Because the packet does not originate from the controller, the malicious relay server is not able to prove its innocence through the rebutting process. Although the attacker can rejoin RS-IoT with a new account, the risk of losing deposit still exists. Also, compared with the random scanning attack, it's very slow to traverse IoT devices on the RS-IoT platform by passively waiting to be commissioned. Finally, attackers get discouraged because this type of attack is not only risky but also inefficient.

\subsection{Fairness Analysis}
Considering there is no trust between relay users and servers, fairness of the service trading platform is required to prevent cheating behaviors of either parties. We enumerate all possible cheating scenarios and show how to deal with them by using our proof-of-delivery scheme.

\subsubsection{Cheating}
First, the relay user has the incentive to cheat by denying that they have already received the relayed packet from the relay server. According to the commitment procedure described in Section \ref{sec:pod}, this can be achieved by sending an incorrect commitment back to the relay server. For this kind of cheating, the relay server cannot verify the commitment $B$ and $B'$ and thus won't reveal the cover stream key $PN$. Without $PN$, the cheating IoT device is not able to extract the desired content, which is equivalent to receiving nothing. Hence, the free ride is impossible due to the packet covering conducted by the relay server.

Second, the relay server has the motivation to reap without sowing. That is, it may  deliver incomplete packets to IoT devices to reduce the cost. Since the delivered packet is covered by the cover stream, IoT cannot verify its integrity before the relay server reveals the cover stream key $PN$. However, the byte selecting list $Ra'$ is not known to the relay server, and it has no idea about which bytes will be selected for composing the commitment. As a result, when an IoT device generates the commitment $B'$ on the imcompletely delivered packet, the relay server has no method to figure out a $PN$ that can satisfy the relation of $Cover_{PN}(B) = B'$. Then, there will be no way to pass the checking of proof-of-delivery by the smart contact to obtain the service fee.


\subsubsection{Malicious Reporting}
Since joining the system as an IoT devices requires no deposit, attackers who want to destroy the system may register a large amount of IoT device accounts and use them to maliciously report benign relay servers. Although we design the rebutting scheme, this method may still be able to overload relay servers and undermine the system's performance.

We prevent this kind of abuse by utilizing the high cost nature of on-blockchain function calls, which is usually regarded as a problem as we discussed before. According to our description of the reporting process, IoT devices need to pass the whole packet as one of the arguments to the function call which will be very expensive considering the data storage and processing price on Ethereum~\cite{wood2014ethereum}. For normal reporting of real malicious packets, this cost will be made up by the confiscated deposit of the relay server. However, if an IoT device reports benign packet, there will be no make up. So, malicious reporting leads to high cost and is quite uneconomic.


%% file: sections/experiment.tex
\section{Experiment}\label{sec:exp}
To validate the efficiency and usability of RS-IoT, we deploy the proposed RS-IoT on Ethereum rinkeby testnet with solidity script language of version 0.4.21.  After that, we use three Raspberry PIs with Geth (Golang Ethereum Client) and the  web3.py package installed as an emulation of the relay server, IoT, and controller respectively. Each of the raspberry PIs has one account setup in its geth client, and is topped up with test ethers from the rinkeby Faucet. To minimize the execution cost, we avoid using local variables and store all intermediate results in a memory location. We set the commitment length $N$ to 32 bytes to avoid exceeding block gas limit. Based on the cost of gas defined in the Ethereum yellow paper~\cite{wood2014ethereum}, we can accurately evaluate the execution cost of every function that we use. Also, with the gas price set to 2 Gwei (1 Ether equals to $1*E^9$ Gwei) and an Ether price of \$135, we can convert the execution cost to USD as listed in Table~\ref{tab:cost}.



\begin{table}[htbp]\caption{Contracts Execution Cost.}
	\begin{center}
		\scalebox{1.0}{
		\begin{tabular}{l c c p{3cm} }
			\textbf{Entity:Function} & \textbf{Cost in Gas} & \textbf{Cost in USD}\\
			\toprule
			\centering
			$D$:register & 47k & 0.012 \\
			$C$:register & 22k & 0.005 \\
			$R$:register & 40k & 0.01 \\
			$D$:service request &1.8k & 0.0003 \\
			$D$:service select &14.3k & 0.003 \\
			$D$:service confirm & 22.8k& 0.005 \\
			$D$:commitment & 175k & 0.046 \\
			$C$:commitment & 151k & 0.039 \\
			$R$:commitment verify &40k& 0.01 \\
			$C,D,R$:decommission& 12k & 0.003 \\
			$D$:execute & 8k & 0.002 \\
			\bottomrule
			Registration Total & 109k & 0.03 \\
			Commission Total & 32.6k & 0.009 \\
			Commit Total &366k & 0.10 \\
			\bottomrule
		\end{tabular}
	}
	\end{center}
	\label{tab:cost}
\end{table}

As we can see, since all the operations of the RS-IoT are asynchronous and none of them exceed the Ethereum gas limit (average 3,000,000 gas per block), there is no limitation on the number of concurrent online relay sessions. From the usage perspective, registration costs 109k Gas in total, which is a one-time cost. Commission/decommission cost 32.6k gas, but it only happens when the IoT device switches to a new relay server. Though the price for commitment is high, asynchronous verification as mentioned in Section~\ref{sec:deliver} makes it unnecessary to be called for each delivered packet. Instead, by caching the commitment transactions, verification can be conducted in arbitrary long packet intervals as long as no disputation emerges. Thus, the cost is amortized as affordable.

\begin{figure}
	\centering
	\includegraphics[width = 0.4\textwidth]{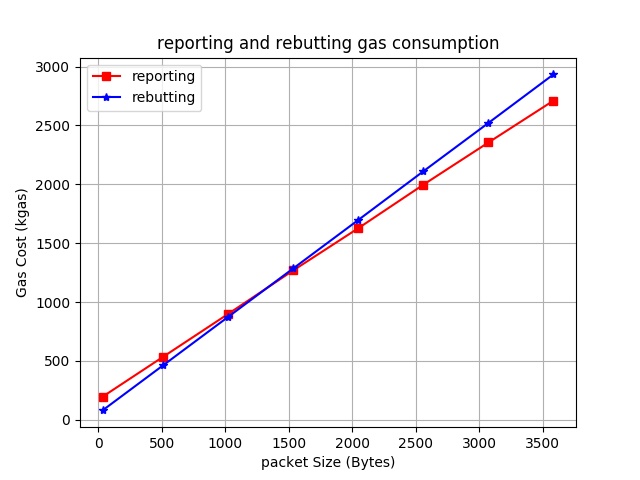}
	\caption{Reporting \& Rebutting Gas Cost.}
	\label{fig:rrgascost}
\end{figure}

The costs of reporting and rebutting scales along with the size of the reported packet are shown in Figure~\ref{fig:rrgascost}. Though the gas limit of one block is about 7 million gas, operations that consume more than 3.5 million gas become difficult to processed. The largest packet size for successful reporting and rebutting in our experiment is 3.5k bytes.


%% file: sections/relatedwork.tex
\section{Related Work}\label{sec:relatedwork}
\subsection{IoT Malware Defense}
After the breakout of Mirai botnet, various defense schemes are proposed. They can be categorized into three types: The first type is the honeypot as in \cite{pa2015iotpot,luo2017iotcandyjar,Antonakakis2017} which is usually for research purpose only. The honeypot is a computer with publicly accessible IP addresses and powerful traffic monitoring and logging systems that imitates normal IoT devices. It functions as a trap to detect botnet scanning traffic and after that entices the loading of executable malware files by pretending to be compromised. In this way, the honeypot operators can acquire samples of malwares at the first time.  Analysis of the sample malware gives security experts clues about the address of the attacking master's command and control server and helps them make security patches. However, because of the lack of reliable firmware update methods, it's hard to push the security patch to vulnerable devices, which undermines the effect of this kind of defense.

The second type is the secure software implementation guideline, and modules like IDS (Intrusion Detection System) designed for IoT devices and IoT connected cloud servers. \cite{barrera2017idiot,payne2018securing} fall into this type.  \cite{barrera2017idiot} proposes to enforce security policies on IoT devices to filter out abnormal or unnecessary packets and \cite{payne2018securing} offers some guidelines from perspectives of network configuration and device deployment. Both of these works require assumptions about the robustness of deployed softwares, which does not always hold.

The last type of defense as in \cite{cao2017hey,lauria2017footprint,goodin2017brickerbot} is actually inspired by Mirai botnet malware, which involves discovering vulnerable IoT devices by random scanning. Upon discovering new targets, they either dig into the system to expel the hidden malware or report the vulnerable device to authorities and the owner. However, this type of defense is limited on its usability that there is normally no reliable way to notify the owner for discovered vulnerabilities. Purifying the IoT device without getting the owner's approval causes legal concern.

\subsection{Blockchain based IoT Service}
Though blockchain has been proved as a successful technology for years, the high cost of transaction fee and limited network throughput have long become the impediment of adopting it for IoT. However, there are quite some works trying to find workaround ways to integrate the power of blockchain into IoT systems. In \cite{andersen2017wave} smart contract is used as an IoT authorization platform where the smart contract only serves as an online distributed ledger to record permission changes. In \cite{hardjono2016cloud}, blockchain is used for secure device commissioning and data privacy protecting with the help of secure hardware. In \cite{dorri2017blockchain}, self-mined private blockchain is deployed in home network as a ledger of accessing privilege and access control policies. However, none of these works solve the problem of IoT malware propagation.

Although the idea of imposing economic incentives and penalties has already been proposed in Enigma~\cite{zyskind2015enigma}, it only provides a theoretical model for secure multi-party computation rather than a usable guideline for IoT and blockchain integration. Enigma conduct most of its operation on the blockchain which is inevitably expensive. In comparison, our proof-of-delivery scheme can run off-chain completely as long as no dispute happens. Other related works that utilizing economic incentives such as NameCoin~\cite{loibl2014namecoin} and FileCoin~\cite{benet2018filecoin} also use the economic incentives, but they are applied on specific tasks that involves low frequency of on-chain operation and do not propose the framework of off-chain verfification.

%% file: sections/conclusion.tex
\section{Discussion}\label{sec:dis}
\paragraph{Device endpoints}
Blockchain data is increasing infinitely, it is unrealistic to let each IoT devices to access the public blockchain with their own client. In real-world deployment, the blockchain client can be hosted on a local edge server (e.g., a home router, an IoT device hub) which has higher computing power and sufficient storage space to host an Ethereum Geth client~\cite{geth}. The edge server opens the Remote Procedure Call (RPC) interface and serves as a proxy for IoT devices to query the blockchain's content and broadcast transactions. As a result, resource limited IoT devices do not need to store any blockchain data except their own private keys for signing transactions. As the edge server is owned and operated by the same owner, it is trusted by the all other IoT devices with in the same household's network. This architecture has already been widely adopted by many other research works and is proved to be more efficient than the light Geth client~\cite{lightgeth}.

\paragraph{Account compromising attacks}
Since each IoT devices need to setup its own cryptocurrency wallet for paying the service fee, malicious relay servers have the motivation to steal their customer IoT devices' account. Malicious servers may launch attacks in the wish of retrieving victim devices' blockchain private key. However, in reality, this type of attack can be easily defeated by allowing deposing service fees with any other blockchain account (e.g., users' personal accounts that are securely protected). There is no need to check the eligibility of making deposits because malicious depositing makes no sense. As a result, there will only be very few amount of cryptocurrency in  IoT devices' accounts for paying transaction fees. As a result, earnings of this kind of attacks are very limited compared to the high risk of being reported and losing a large amount of the deposit.


\paragraph{Service Quality}
Different relay servers may provide services with different qualities. For example, a relay server with more powerful hardware and higher Internet bandwidth could forward packets with smaller latency. To help relay service users to choose better service providers, the service rating function can be added to the current smart contract. For a specific relay server, its service users that have use it to forward a higher number of packets than a pre-defined threshold have the eligibility to rate its service quality. The smart contract will record its total number of ratings and the average rating score along with its record in the \textit{serverInfo} table. Hence, on receiving quotes from relay servers, IoT devices have additional information to find a most attactive one with balanced price and service quality.


\paragraph{Amount of Service Fee and Deposit}
This paper concentrates on demonstrating the framework of IoT and blockchain integration. Finding the optimal amount of service fee and deposit requires economic theories that is out of the scope of this paper. Intuitively, the required deposit for a specific relay server can be dynamically tuned according to the number of its users. When the number of users reaches a higher lever, the relay server should invest additional deposit accordingly.

\section{Conclusion}\label{sec:con}
In this paper, we presented a general framework for efficient blockchian and IoT service integration. To solve high cost and overhead of on-blockchain operations, we proposed the distributed architecture to host IoT services on third-part servers and use the blockchain and the smart contract as a naturally trusted authority to enforce the fairness and punish attackers. We applied this architecture on the task of IoT remote accessing and designed RS-IoT, a blockchain-assisted distributed relay sharing system. RS-IoT provided secure and robust relay services for IoT users to access their IoT devices which are behind the network address translation (NAT). We utilized ``\emph{an economic approach to cyber security}'' to deter malicious relay servers and achived it with our novel proof-of-delivery mechanism. By verifying proofs off-chain, the costs and throughput issues of blockchain are overcome. We demonstrated the cost efficiency of our design with our prototype implementation on the Ethereum testnet.
